\documentclass[
superscriptaddress,
preprint,
nofootinbib,
amsmath,amssymb,
aps,
pra,
floatfix,
longbibliography
]{revtex4-1}
\usepackage{graphicx}
\usepackage{subfigure}
\usepackage{physics}
\usepackage{placeins}
\usepackage{appendix}
\usepackage{float}
\usepackage{graphicx}
\usepackage{dcolumn}
\usepackage{bm}
\usepackage{booktabs}
\usepackage{hyperref}
\usepackage{xcolor}
\hypersetup{
    colorlinks,
    linkcolor={blue!70!black},
    citecolor={blue!70!black},
    urlcolor={blue!70!black}
}

\begin{document}

\title{Optical cycling, radiative deflection and laser cooling of \\* barium monohydride (\textsuperscript{138}Ba\textsuperscript{1}H)}

\author{Rees L. McNally}
\email{rm3334@columbia.edu}
\affiliation{Department of Physics, Columbia University, 538 West 120th Street, New York, NY 10027-5255, USA}

\author{Ivan Kozyryev}
\affiliation{Department of Physics, Columbia University, 538 West 120th Street, New York, NY 10027-5255, USA}

\author{Sebastian Vazquez-Carson}
\affiliation{Department of Physics, Columbia University, 538 West 120th Street, New York, NY 10027-5255, USA}

\author{Konrad Wenz}
\affiliation{Department of Physics, Columbia University, 538 West 120th Street, New York, NY 10027-5255, USA}

\author{Tianli Wang}
\affiliation{Department of Physics, Columbia University, 538 West 120th Street, New York, NY 10027-5255, USA}
\address{Present address:  Department of Physics, Ludwig Maximilian University of Munich, Geschwister-Scholl-Platz 1, 80539 M\"{u}nchen, Germany}

\author{Tanya Zelevinsky}
\email{tanya.zelevinsky@columbia.edu}
\affiliation{Department of Physics, Columbia University, 538 West 120th Street, New York, NY 10027-5255, USA}

\date{\today}

\begin{abstract}
We present the first experimental demonstration of radiation pressure force deflection and direct laser cooling for barium monohydride (BaH) molecules resulting from multiple photon scattering. Despite a small recoil velocity (2.7 mm/s) and a long excited state lifetime (137 ns), we use 1060 nm laser light exciting the $X\rightarrow A$ electronic transition of BaH to deflect a cryogenic buffer-gas beam and reduce its transverse velocity spread. Multiple experimental methods are employed to characterize the optical cycling dynamics and benchmark theoretical estimates based on rate equation models as well as solutions of the Lindblad master equation for the complete multilevel system. Broader implications for laser cooling and magneto-optical trapping of heavy-metal-containing molecules with narrow transition linewidths are presented. Our results pave the way for producing a new class of ultracold molecules -- alkaline earth monohydrides -- via direct laser cooling and trapping, opening the door to realizing a new method for delivering ultracold hydrogen atoms (Lane 2015 \textit{Phys. Rev. A} 92, 022511). 
\end{abstract}
\maketitle

\section{Introduction \label{sec:Intro}}
Various approaches have been developed for controlling atomic motion \cite{greytak1998bec,doret2009buffer,raizen2009comprehensive}. Radiation pressure force and associated laser cooling and trapping methods in particular have revolutionized atomic physics \cite{chu1998nobel,phillips1998nobel} and have since become invaluable tools for modern quantum science and engineering
\cite{weiss2017quantum}. While primarily relying on rigorous angular momentum selection rules present only in a handful of atoms, laser-cooled atomic samples have provided an extremely fruitful experimental platform for probing important condensed matter models \cite{gross2017quantum}, studying exotic phases of matter \cite{lewenstein2007ultracold}, and developing new quantum sensors and ultra-precise clocks \cite{bongs2019taking}. Already for diatomic molecules, the increased complexity of internal structure combined with various couplings between electronic, vibrational, and rotational dynamics preclude the existence of a pure two-level substructure required for repeated photon scattering \cite{tarbutt2018laser}. 

In order to extend laser cooling to molecules, the use of 
quasi-closed cycling
rovibrational transitions together with extra vibrational repumping lasers has been proposed \cite{stuhl2008magneto,di2004laser}. Spurred by such conceptual advances as well as by the latest developments in continuous-wave laser technology, major inroads in direct laser cooling of diatomic and even polyatomic molecules have been made in the last decade \cite{McCarron2018laser}, culminating with the demonstration of three-dimensional magneto-optical trapping at sub-millikelvin temperatures for several molecular species \cite{norrgard2016submillikelvin,truppe2017molecules,anderegg2017radio,collopy20183d}. Full control of molecular degrees of freedom at ultracold temperatures provides unique opportunities for realizing new applications, including quantum simulations of strongly interacting many-body systems \cite{hazzard2014many}, low-energy precision searches for physics beyond the Standard Model \cite{demille2015diatomic,kondov2019molecular}, experimental tests of fundamental chemical processes \cite{mcdonald2016photodissociation}, and implementation of quantum computation protocols \cite{blackmore2018ultracold,sawant2020qudits}. However, before these and many other possible applications \cite{carr2009cold} can be explored to the fullest extent, robust methods for generation of ultracold molecular samples with diverse properties and constituents must be realized and carefully characterized. A critical step towards this goal for each new molecular species is the demonstration of sustained optical cycling without considerable loss to unaddressed dark states \cite{albrecht2020buffer}, and control over molecular motion using laser light \cite{chen2017radiative}. 

Here, we demonstrate such optical cycling and radiation pressure force milestones for diatomic barium monohydride (BaH), the first metal monohydride molecule to be directly laser cooled. Experimental techniques for achieving ultracold heavy-metal monohydrides have potential to open applications in probing fundamental symmetry-violating interactions \cite{fazil2019rah,Berger2019Systematic} as well developing a novel method for indirect production of ultracold atomic hydrogen via precision photodissociation \cite{lane2015ultracold}. Our results are provided in order of increasing implementation complexity, presenting a roadmap for future experiments: i) detailed analysis of optical cycling ($\sim 10-80$ photons) using both depletion and deflection of a cryogenic beam, ii) laser compression of a molecular beam along one transverse dimension ($\gtrsim 100$ photons), and iii) sustained optical cycling in a longitudinal slowing configuration ($\gtrsim 1,000$ photons). Our observations clearly show spatial manipulation of the BaH molecular beam, and demonstrate understanding of the optical cycling process out to several thousand photons. In addition to achieving the first optical manipulation of a heavy-metal monohydride molecule, we provide theoretical modeling of the optical cycling and magneto-optical trapping process for BaH. Taken together, our experimental and theoretical results provide important insights into the challenges of laser cooling and trapping for heavy molecules with reasonably narrow transition linewidths ($\Gamma_{\rm{nat}}\lesssim 2\pi\times 10^6\,\rm{s^{-1}}$), a molecular class frequently encountered in fundamental physics applications \cite{norrgard2017hyperfine,fazil2019rah,Kudrin2019TlCN}.   

\section{Experimental Apparatus \label{Apparatus}}
Compared to previously laser cooled molecules, which have excited state lifetimes comparable to alkali metals ($\tau_{\rm{sp}}\approx 20-30$ ns), BaH has a much longer excited state lifetime for any of the possible optical cycling transitions ($\tau_{\rm{sp}}>100$ ns). A small photon recoil velocity ($v_{\rm{recoil}}\lesssim 3$ mm/s) together with a low possible scattering rate ($R_{\rm{scat}}\lesssim 1\times 10^6$ s\textsuperscript{-1}) require a use of a slow cryogenic buffer-gas beam \cite{hutzler2012buffer} in order to provide enough molecule-laser interaction time to observe radiation pressure effects. Therefore, each of the following experiments was performed using a cryogenic molecular beam of BaH, whose construction, optimization and operation have been described in our previous publication \cite{iwata2017high}. Because of the small capture velocity for the radiation pressure force $\triangle v_{\rm rad}\sim\Gamma_{\rm{sp}}/k\approx 1$ m/s for BaH, we installed a 5-mm circular collimator
48 cm away from the cryogenic cell aperture to match the transverse velocity spread of the molecular beam to $\triangle v_{\rm{rad}}$.  

In order to allow for a sufficient atom-molecule interaction time to observe transverse laser cooling of the BaH cryogenic beam, we used 15-cm long vacuum viewport windows anti-reflection coated for a reflectivity of $r_{\rm{window}}<0.5\%$ per surface and out-of-vacuum mirrors with dielectric high-reflectivity coatings with $r_{\rm{mirror}}>99.8\%$. Therefore, with 10 round-trip passes the light intensity drops to about $80\%$ and with 20 passes to $60\%$, which presents a
challenge for maintaining a uniform scattering rate across the entire region, requiring a higher incident laser power. For the radiative deflection experiments, we used hollow rooftop mirrors (HRM) out of vacuum since the photon momentum kicks have to come from a single direction. Gold mirror coatings provided $r_{\rm{HRM}}\approx 97\%$ which limited the number of passes with sufficiently high laser intensity to achieve efficient photon scattering.   

\section{Molecular Structure and Optical Cycling Scheme \label{sec:Mol-struct}}
There are a number of important differences between the relevant structure of BaH and that of other previously laser cooled diatomic and triatomic molecules. Except for YO \cite{hummon2013YO}, all of the diatomic and triatomic molecules laser cooled thus far consisted of an alkaline-earth-metal like atom (Ca, Sr or Yb) ionically and monovalently bonded to an electronegative fluorine (F) \cite{shuman2010laser,zhelyazkova2014laser,lim2018laser} or hydroxyl (OH) \cite{kozyryev2017sisyphus,augenbraun2020laser,baum2020CaOH} ligand. Such ligands with a moderate electron-withdrawing capability lead to a strong localization of a single unpaired electron on the metal atom with electronic excitations resembling those of an ionized alkaline earth metal atom M$^+$ \cite{Ivanov2019Rational}. However, because the partial charge of the metal increases with the electron affinity of the ligand \cite{Bernath1991}, upon electronic excitation the M-L bond length change is largest for monohydrides and decreases for MF and MOH compounds \cite{Ivanov2019Rational}. More covalent nature of the metal-ligand bond for hydrides has been further confirmed by the estimated values of the quantum defect across a range of ligands attached to the same alkaline earth metal \cite{Ivanov2019Rational}. While previous theoretical studies \cite{li2019emulating,klos2020prospects} have indicated that a large number of strongly ionically bonded compounds (e.g. MF, MOH and MOR where R is a functional group) are well suited for laser cooling, other molecules with different geometries (MCH$_3$) and constituents (MSH) have more covalent nature of M-L bonds \cite{Bernath1997}, with BaH potentially serving as an important stepping stone for understanding optical cycling and laser cooling in covalently-bonded systems \cite{augenbraun2020ATMs}. 

\begin{figure}[h]
\centering
\includegraphics[width=.7 \textwidth]{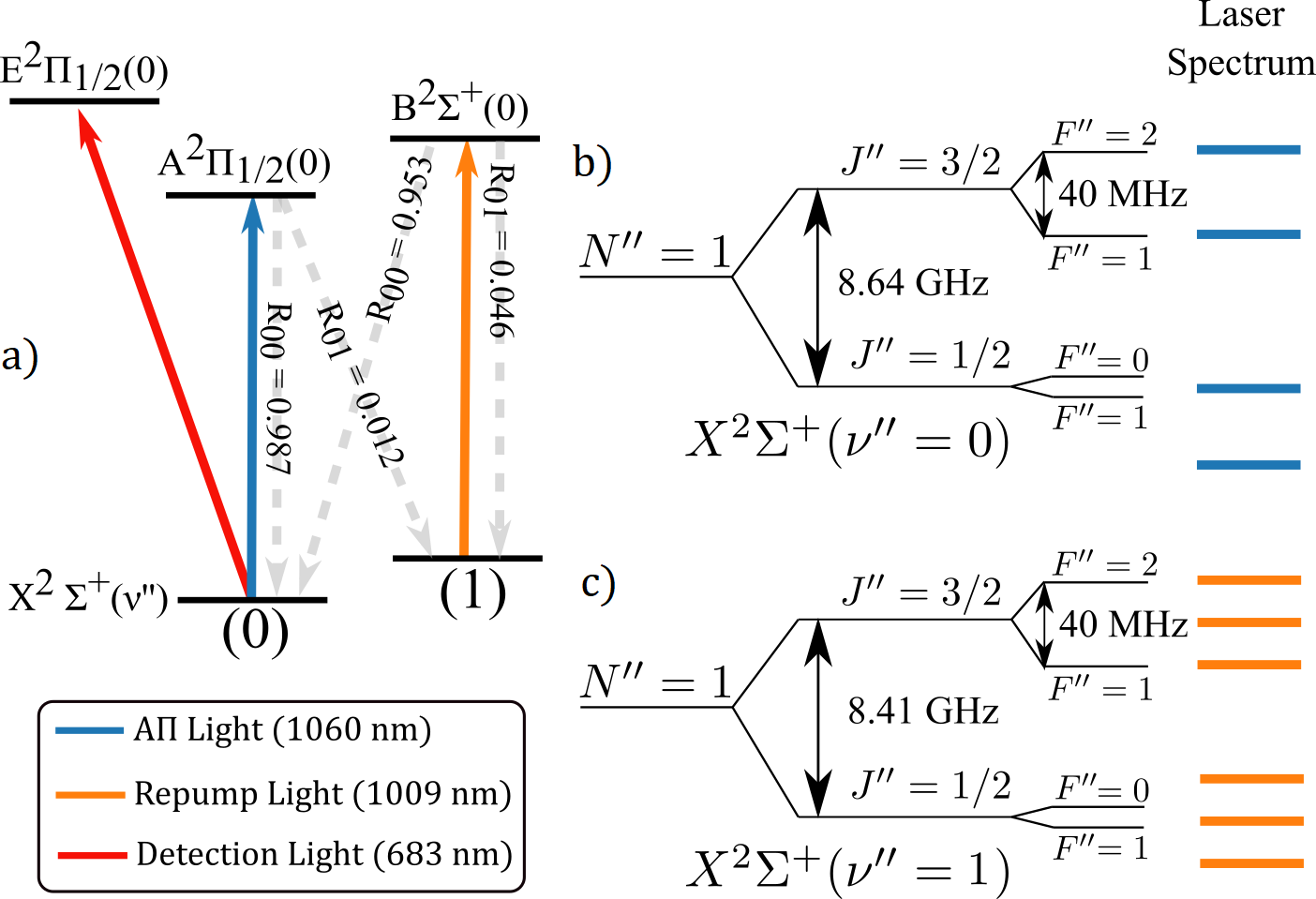}
\caption{(a) BaH electronic level structure as well as laser wavelengths used for optical cycling (1060 nm), vibrational repumping (1009 nm), and non-cycling beam detection (683 nm). Numbers in parentheses $\left(v\right)$ indicate vibrational quantum numbers. The branching ratios are the combined theoretical and experimental results from our previous work \cite{moore2019assignment}. Spin-rotation and hyperfine structure for the $N''=1$ rotational levels of the ground electronic state together with the laser sidebands used for the $\left(v''=0\right)$ (b) and $\left(v''=1\right)$ (c) vibrational states.}
\label{fig:level-structure}
\end{figure}

One of the unique aspects of the BaH level structure 
is that it supports optical cycling on three different electronic transitions within a technically convenient near-infrared wavelength regime but with distinct laser cooling characteristics: i) $X^2\Sigma^+\leftrightarrow A^2\Pi_{1/2}$ at 1061 nm ($\tau_{\rm{sp}}\approx140$ ns, Doppler cooling limit $T_D\approx30\,\rm{\mu K}$), ii) $X^2\Sigma^+\leftrightarrow B^2\Sigma^+$ at 905 nm ($\tau_{\rm{sp}}\approx130$ ns, $T_D\approx30\,\rm{\mu K}$), and iii) $X^2\Sigma^+\leftrightarrow H^2\triangle_{3/2}$ at 1110 nm ($\tau_{\rm{sp}}\approx10\,\rm{\mu s}$, $T_D\approx0.4\,\rm{\mu K}$) electronic transitions \cite{moore2019assignment}. Due to the highly diagonal nature of the Frank-Condon factor (FCF) matrix ($\mathcal{F}_{v''v'}$) for the $X^2\Sigma^+\leftarrow A^2\Pi_{1/2}$ electronic decay\footnote{Following the established convention in the field of molecular spectroscopy we mark quantum numbers with double (single) primes to refer to the electronic ground (excited) state, correspondingly.} ($\mathcal{F}_{00}>0.987$ \cite{moore2019assignment}), we choose to use this transition for optical cycling, and repump molecules from excited vibrational levels $v''>0$ back into the cycle with an off-diagonal transition through the $B^2\Sigma^+$ excited electronic state (Fig. \ref{fig:level-structure}(a)). By separating the cycling and repumping transitions (\textit{i.e.} they are not directly coupled with laser light via a common vibronic manifold), we increase the maximum achievable scattering rate by a factor of 1.75. 

Following previous experiments \cite{shuman2009radiative,McCarron2018laser}, we drive all optical transitions from the first excited rotational state $N''=1$ of the electronic ground state $X$ to the ground rotational level $N'=0$ in each excited electronic state ($A$ or $B$) in order to ensure rotational closure (\textit{i.e.} $N''=1\leftrightarrow N'=0$). Parity as well as angular momentum selection rules for the electric dipole allowed transitions ensure that molecules decay back to the $N''=1$ rotational manifold of states \cite{stuhl2008magneto}, forming a quasi-closed cycling transition required for repeated photon scattering. Optical cycling on a $\Delta N = -1$ transition has the additional requirement that dark states need to be destabilized \cite{shuman2009radiative}, which we achieved by the addition of a static magnetic field. 

Because vibrational, rotational, and spin-rotational molecular constants in $^2\Sigma^+$ electronic states scale as $\omega_{\rm{vib}}\propto \mu_{\rm{red}}^{-1/2}$, $B_{\rm{rot}}\propto \mu_{\rm{red}}^{-1}$ and $\gamma_{\rm{SR}}\propto \mu_{\rm{red}}^{-1}$  with reduced molecular mass $\mu_{\rm{red}}$ \cite{veseth1971spin}, correspondingly large energy spacings in BaH (compared to MF or MOH molecules) result in additional challenges for optical cycling and laser cooling. Because both $J$-sublevels of the spin-rotation splittings in the $N''=1$ states ($1.5\gamma_{\mathrm{SR},v''=0}=8.64$ GHz \cite{tarallo2016bah} and $1.5\gamma_{\mathrm{SR},v''=1}=8.41$ GHz) cannot be easily addressed with the same light source, we use two separate external cavity diode lasers (ECDLs) to excite $J''=1/2$ and $J''=3/2$ states (Fig. \ref{fig:level-structure}(b) and \ref{fig:level-structure}(c)), with $1.5\gamma_{\mathrm{SR},v''}$ offsets.
The two pairs of resulting ECDLs
are co-aligned with matching linear polarization to seed two tapered amplifiers producing $\sim100$ mW of 1060 nm light and $\sim30$ mW of 1009 nm light in the beam cooling region. For the $\left(0,0\right)X^2\Sigma^+\rightarrow A^2\Pi_{1/2}$ transition\footnote{Following the convention in the spectroscopic literature for diatomic molecules \cite{steimle2004CaH}, we use the $\left(v'',v'\right)X\rightarrow A$ notation to indicate the change in the vibrational quantum number from $v''$ in the $X$ electronic state to $v'$ in the $A$ state.}, the hyperfine sidebands are generated using a 40 MHz AOM, which leads to one additional off-resonant sideband for the $J''=1/2$ hyperfine states (Fig. \ref{fig:level-structure}(b)). For the $\left(0,1\right)X^2\Sigma^+\rightarrow B^2\Sigma^+$ vibrational repumper, the hyperfine structure is addressed using a 20 MHz EOM driven with a modulation depth set to optimize the power in the $\pm$ 1st order. Each laser frequency is stabilized via referencing to a HighFinesse WS7 wavemeter, which provides a short term ($\sim1$ s) instability of  $\sim1$ MHz and a longer term ($\sim5$ h) instability of $\lesssim5$ MHz, consistent with performance achieved in other experiments \cite{couturier2018laser}. Frequency stability of the reference wavemeter was additionally verified by monitoring the wavelength of a frequency-comb stabilized ECDL over the course of an hour and by daily calibration of the wavemeter with a frequency-stabilized HeNe laser.

We use the $X^2\Sigma^+ \rightarrow E^2\Pi_{1/2}$ excitation at 683 nm to collect the time-of-flight (ToF) data and spatial distribution images of the BaH molecular beam because this wavelength matches the peak sensitivity of EMCCD and PMT detectors used to detect these molecules. For this transition we also use two lasers, with the $J=3/2$ hyperfine splitting addressed with a 40 MHz AOM. Both E$^2\Pi_{1/2}$ lasers are broadened with a 3 MHz EOM, to ensure that they interact with all velocity classes and that the fluorescence signal accurately represents the total $v''=0$ population regardless of the specific hyperfine distribution of the molecular ensemble when it reaches the detection region.

\section{Optical Cycling \label{sec:Optical-cycling}}

\begin{figure}[h]
\centering
\includegraphics[width=.8 \textwidth]{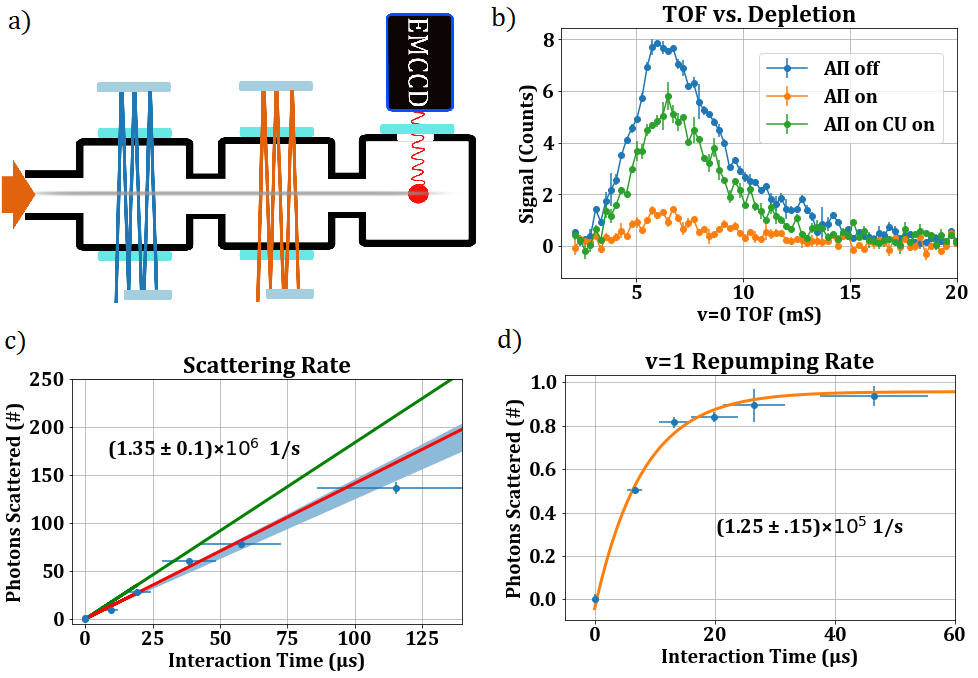}
\caption{(a) Diagram of the experimental setup showing the spatially separated depletion and clean-up (CU) regions, followed by $\left(0,0\right)X^2\Sigma^+\rightarrow E^2\Pi_{1/2}$ fluorescence detection. b) Example time-of-flight signals for the unperturbed (A$\Pi$ off), depleted (A$\Pi$ on), and vibrationally repumped (A$\Pi$ on and CU on) molecular beam signals. c) Scattering rate for the $\left(0,0\right)X^2\Sigma^+\rightarrow A^2\Pi_{1/2}$ transition, based on depletion of $v''=0$ population
as a function of interaction time. Blue points and band represent the data and the $1\sigma$ uncertainty from a linear fit.
Green (slope $1.84\times 10^6$ s\textsuperscript{-1}) and red (slope $1.41\times 10^6$ s\textsuperscript{-1}) lines show the theoretical estimates obtained from Eq. \ref{eq:R-scatter-max} and from a full numerical simulation of our system using the Lindblad master equation (Appendix \ref{sec:OBE}). d) Estimation of scattering rate for the repumping transition based on repopulation of the $v''=0$ state. We find a rate of $1.25 \times 10^5$ photons$/$s. Note, that an offset to the fit model is allowed to account for imperfect alignment of the CU laser and the depletion laser.}
\label{fig:Photon-cycling}
\end{figure}

The first step in achieving laser control and cooling of molecular motion is to establish repeated scattering of photons (optical cycling) and characterize dominant loss channels. As discussed in Sec. \ref{sec:Mol-struct}, our detection scheme relies on a non-cycling transition at 683 nm, necessitating a different approach to characterizing the photon scattering dynamics for the main laser cooling transition at 1060 nm. The experimental setup used for characterizing the $\left(0,0\right)X^2\Sigma^+\leftrightarrow A^2\Pi_{1/2}$ scattering rate is shown in Fig. \ref{fig:Photon-cycling}(a): the  $\left(0,1\right)X^2\Sigma^+\rightarrow B^2\Sigma^+$ repumping light (orange) was blocked, and the number of passes of the $\left(0,0\right)X^2\Sigma^+\rightarrow A^2\Pi_{1/2}$
 laser (blue) was varied. The 1060 nm laser beam was alternated between ``on" and ``off"
to account for any drift in the molecular beam yield,
and each data point is the average of 200 molecular beam pulses. Accounting for the beam forward velocity ($160 \pm 40$ m/s \cite{iwata2017high}) and the measured diameter of the laser beam ($1.5 \pm 0.1$ mm), we can convert the number of passes to the molecule-light interaction time. As seen from the error bars in Fig. \ref{fig:Photon-cycling}(c,d), this conversion is the dominant source of uncertainty,
primarily because of the substantial spread in the beam forward velocity. We can estimate the number of photons scattered $\left(N_{\rm{scat}}\right)$ based on the fraction of molecules that remain in ground vibrational state ($P_{v=0}$), and the known diagonal FCF $\mathcal{F}_{00}$ from previous measurements \cite{moore2018quantitative,moore2019assignment}. Figure \ref{fig:Photon-cycling}(b) provides a representative ToF data for an unperturbed BaH beam (blue) and with the $\left(0,0\right)X^2\Sigma^+\rightarrow A^2\Pi_{1/2}$ cycling laser on (orange) resulting in 15\% of the molecules remaining in $v''=0$ at the detection region. We found no significant dependence of the scattering rate on an applied magnetic field used to destabilize the dark states, most likely because residual field in the interaction region of a few Gauss was sufficient to cause a dark state precession rate comparable to the excitation rate ($\sim 10^6$ s\textsuperscript{-1}).

Following Di Rosa \cite{di2004laser}, we model the repeated 
spontaneous emission events by a molecule as a 
Bernoulli sequence with probability $p=1-\mathcal{F}_{00}$ that decay will result in populating an excited vibrational level $v''>0$. The probability that a molecule initially in the vibrational ground state will still be in $v''=0$ after scattering $N_{\rm{scat}}$ photons is given by:
\begin{equation}
 P_{v''=0} = \left(\mathcal{F}_{00}\right)^{N_{\rm{scat}}}.
\end{equation}
Therefore, we can convert the fraction $P_{v''=0}$ into the number of scattered photons for the remaining molecules:
\begin{equation}
 N_{\rm{scat}} = \frac{\log P_{v''=0}}{\log \mathcal{F}_{00}}.
 \label{eq:N-scatter}
\end{equation}
The expectation value of $N_{\rm{scat}}$ for a molecular ensemble 
can be estimated by modeling the photon scattering process before the molecule is optically pumped into $v''=1$ as a geometric distribution with the expected value of
\begin{equation}
\langle N_{\rm{scat}} \rangle = \frac{1}{1-\mathcal{F}_{00}} \approx 80,
 \label{eq:N-avg}
\end{equation}
and the standard deviation in $N_{\rm{scat}}$ of
\begin{equation}
\sigma_{N_{\rm{scat}}} = \frac{\sqrt{\mathcal{F}_{00}}}{1-\mathcal{F}_{00}} \approx 80.
 \label{eq:N-std}
\end{equation}
Equation \ref{eq:N-scatter} allows us to estimate the photon scattering rate for the molecules remaining in $v''=0$ as a function of the laser interaction time, $N_{\rm{scat}}=R_{\rm{scat}}t_{\rm{int}}$. However, as shown in equations \ref{eq:N-avg} and \ref{eq:N-std}, for any specific molecule there is a wide range of actual number of scattering events, meaning this setup would not be efficient for optical manipulation. As shown in Fig. \ref{fig:Photon-cycling}(c), a constant scattering rate of $1.4(1)\times 10^6/$s can be maintained, which is $\sim80\%$ of the expected maximum scattering rate $R_{\rm{scat,max}}$ based on the ground and excited state multiplicities,
\begin{equation}
 R_{\rm{scat,max}}=\frac{1}{\tau_{\rm{sp}}} \frac{n_e}{n_e+n_g} = 1.8\times10^6/\mathrm{s},
  \label{eq:R-scatter-max}
\end{equation}
where $\tau_{\rm{sp}}=136.5$ ns \cite{moore2019assignment} is the spontaneous excited state lifetime and $n_g$ ($n_e$) is the number of ground (excited) $m_F$ magnetic sublevels. While Eq. (\ref{eq:R-scatter-max}) provides a useful way to approximate the maximum possible scattering rate for molecules, our estimates for the achievable scattering rate in the experiment using both a multilevel rate equation model (Fig. \ref{fig:Predicted-cycling}) as well as an optimized numerical simulation of the full system using the Lindblad master equation\footnote{The provided scattering rate $R_{\rm{OBE}}$ is the result of an optimization performed for the laser power, detunings and polarizations, and the applied magnetic field, with details provided in Appendix \ref{sec:OBE}.} predict $R_{\rm{OBE}}\approx 1.4\times 10^6$ s\textsuperscript{-1}. The data presented in Fig. \ref{fig:Photon-cycling}(c) shows that we achieve $R_{\rm{scat}}\approx 0.96R_{\rm{OBE}}$ in this experimental configuration with maximum interaction time $t_{\rm{int}}\approx 900\tau_{\rm{sp}}$. 

Our measurement of the scattering rate relies on $\left(1,0\right)X^2\Sigma^+\leftarrow A^2\Pi_{1/2}$
 being the dominant loss mechanism out of the quasi-cycling transition. As shown in Fig. \ref{fig:Photon-cycling}(b), with an addition of the  $\left(0,1\right)X^2\Sigma^+\rightarrow B^2\Sigma^+$ repumping laser in the ``clean-up'' region, we return most of the molecules (green curve) back into the ground vibrational level. The $B^2\Sigma^+\left(v'=0\right)$ state has a good Franck-Condon overlap with $X^2\Sigma^+\left(v''=0\right)$ ($\mathcal{F}_{00}$ = 0.953 \cite{moore2019assignment}) so molecules excited to this state decay to the desired ground state $\ket{v''=0,N''=1}$ with a $95\%$ probability. In order to determine the photon scattering rate for the  $\left(0,1\right)X^2\Sigma^+\rightarrow B^2\Sigma^+$ repumping transition, we begin by depleting $v''=0$ on the main cycling transition $\left(0,0\right)X^2\Sigma^+\rightarrow A^2\Pi_{1/2}$
  at the maximum interaction time. Then in a separate region (Fig. \ref{fig:Photon-cycling}(a)) we apply the repumping light, while varying the number of passes through the molecular beam. Because $\mathcal{F}_{00}\approx 1$ for the 
 $\left(0,0\right)X^2\Sigma^+\leftarrow B^2\Sigma^+$ transition, it will only take $1/\mathcal{F}_{00}\approx$1 photon scattered from the repumping laser to optically pump a molecule back to $v''=0$. As shown in Fig. \ref{fig:Photon-cycling}(d), we model the interaction time required for scattering one photon in the CU region as an exponential distribution 
 with a cumulative distribution function given as $N_{\rm{scat}}=1-\exp{-R_{\rm{scat}}t_{\rm{int}}}$, which becomes $N_{\rm{scat}}\approx R_{\rm{scat}}t_{\rm{int}}$ for $t_{\rm{int}}\ll 1/R_{\rm{scat}}$ and allows us to extract a scattering rate of $1.3(2)\times 10^5$ s\textsuperscript{-1}. Since $R_{\rm{scat}}\propto \sigma_{\rm{abs}}I_0$ and the resonant absorption cross section depends on the corresponding FCF $\sigma_{\rm{abs}}\propto \mathcal{F}_{v''v'}$, the scattering rate will be lower for the off-diagonal transition for a given laser intensity $I_0$. However, using the experimentally measured $R_{\rm{scat}}$ for the main cycling $\left(0,0\right)X^2\Sigma^+\rightarrow A^2\Pi_{1/2}$
 excitation, together with the estimate of the off-diagonal FCF $\mathcal{F}_{01}\approx 0.012$, we determine that the rate of optical pumping into the excited vibrational level $v''=1$ in the optical cycling region ($\sim1.7\times 10^4$ s$^{-1}$) is a factor of $7$ less than
our measured repumping rate, indicating that there is sufficient repumping laser intensity to rapidly return the molecules into the optical cycle. 

\section{Radiative Deflection \label{sec:Radiative-deflection}}
\begin{figure}[h]
\centering
\includegraphics[width=.8 \textwidth]{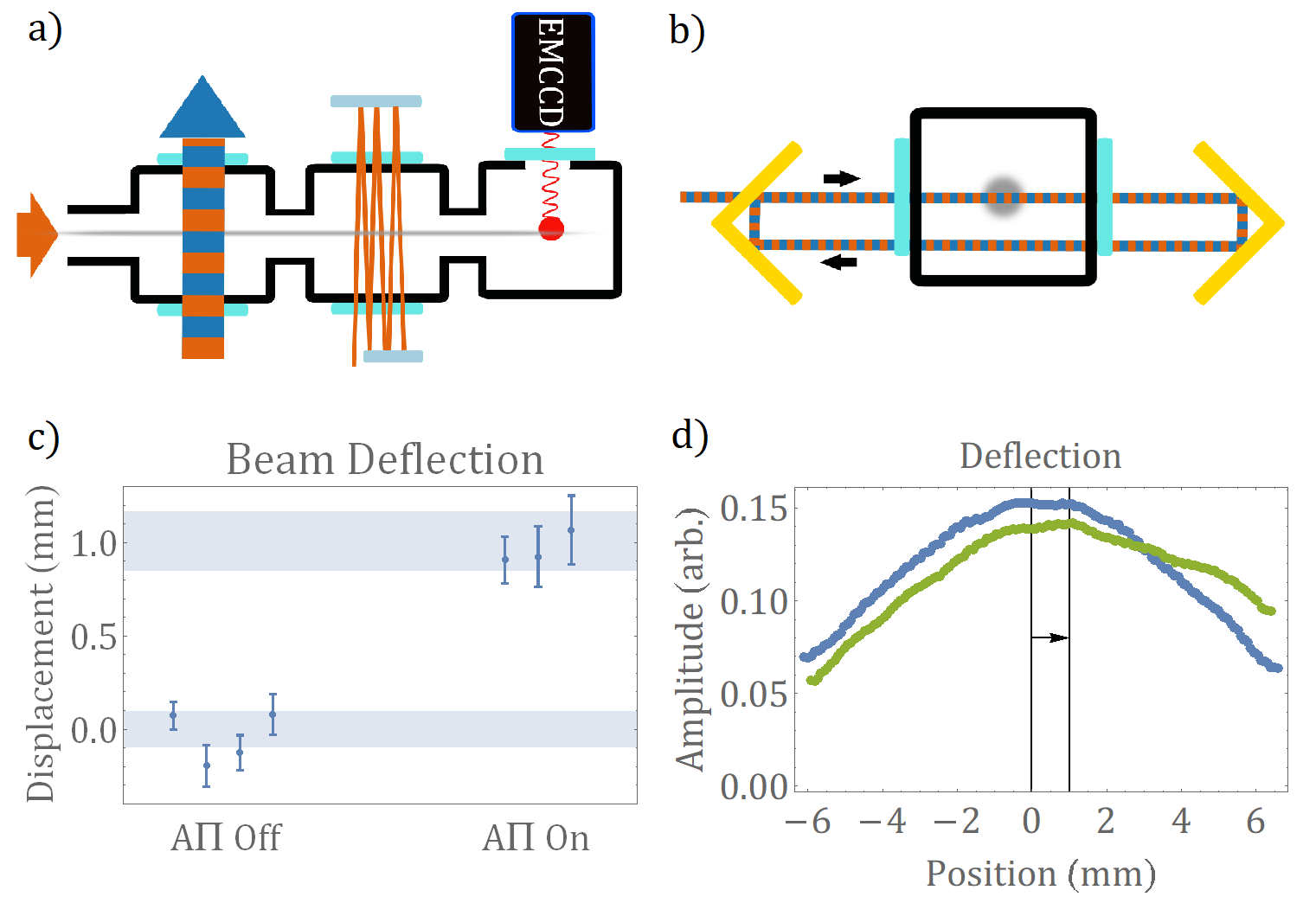}
\caption{a) Schematic of the beam deflection experiment. In the first interaction region, the molecular beam interacts with both the cooling and the repumping light, with each pass coming from the same direction. The molecules then enter a clean-up region where they are pumped back to $v''=0$, and then a detection region 75 cm away where their spatial location is imaged on an EMCCD camera. b) Diagram of the mirror prisms that allow the laser light to interact with the molecular beam while always traveling in the same direction. c) Measured deflection of the molecular beam, where the deflection light is either blocked or applied, showing the statistical uncertainty. Each point is the average of $\sim200$ images. c) Taking 1D cuts of each beam image, combining them, and applying a smoothing average filter allows us to visualize the 1 mm beam deflection.}
\label{fig:Deflection}
\end{figure}

The depletion-based scattering measurements described in Sec. \ref{sec:Optical-cycling} provide strong evidence that we maintain a sufficiently high scattering rate in the optical cycling region to pump most of the molecules from the $v''=0$ vibrational manifold into $v''=1$. Moreover, we achieve a repumping rate
that is significantly higher than the rate of optical pumping into $v''=1$. Therefore, by merging both $\left(0,0\right)X^2\Sigma^+\rightarrow A^2\Pi_{1/2}$ main and 
 $\left(0,1\right)X^2\Sigma^+\rightarrow B^2\Sigma^+$ repumping lasers (Fig. \ref{fig:Deflection}(a)) we expect to deflect the BaH molecular beam using the radiation pressure force. In this experiment, the deflection laser passes through the vacuum chamber perpendicular to the molecular beam and strikes a $90^{\circ}$ mirror prism (a hollow rooftop mirror, HRM) which reflects the light back through the vacuum chamber but displaced downward by $\sim2$ cm, thus traversing below the molecular beam. The light then strikes another $90^{\circ}$ mirror prism that translates the beam upward and redirects it back through the molecules, deflecting the molecular beam in the same direction as the first pass (Fig. \ref{fig:Deflection}(b)). The process is repeated $\gtrsim10$ times to increase the interaction time while maximizing the laser intensity. Since the number of mirror bounces is doubled in order to
propagate the light from a single direction and because the reflectivity $r_{\mathrm{HRM}}<r_{\mathrm{mirror}}$, the effective molecule-light interaction time is shorter than in Sec. \ref{sec:Optical-cycling}.

Figures \ref{fig:Deflection}(c,d) summarize the radiative deflection experiments for BaH. The measured center-of-mass molecular beam deflection is $\sim1$ mm, which is consistent with each molecule scattering $\sim80$ photons at the average rate of $8\times 10^5$ photons$/$s. A number of experimental differences could explain why this scattering rate is $40\%$ slower than that measured via depletion experiments: i) slower than anticipated repumping in the main region, ii) the presence of transverse Doppler shifts,
and iii) the presence of
imperfectly remixed ``dark" states which do not play a dominant role in depletion experiments. Importantly, we do not see an appreciable increase in the width of the molecular beam (Fig. \ref{fig:Deflection}(d)), indicating consistent scattering for each detected molecule.
The scattering rate of $8\times 10^5$ s\textsuperscript{-1} extracted from the deflection of the molecular beam accurately represents the scattering rate that we can achieve and maintain for the entire ensemble of molecules.  



\section{Transverse Laser Cooling \label{sec:Laser-cooling}}
\begin{figure}[h]
\centering
\includegraphics[width=.8 \textwidth]{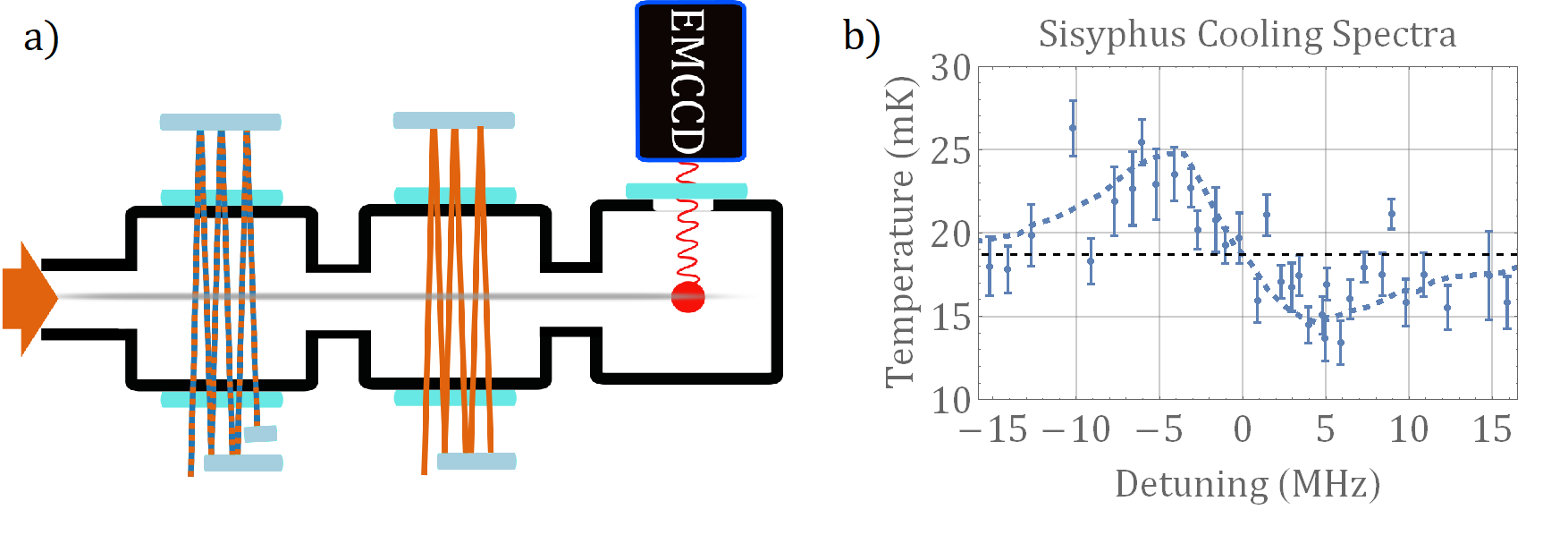}
\caption{a) Experimental diagram showing the standing wave generated by the cooling light that includes both the cycling and repumping lasers. Before molecules enter the detection region, they are optically pumped back to $v''=0$ using the off-diagonal  $\left(0,1\right)X^2\Sigma^+\rightarrow B^2\Sigma^+$ excitation. b) Effective transverse temperature as a function of the common detuning of both spin-rotation components of the $\left(0,0\right)X^2\Sigma^+\rightarrow A^2\Pi_{1/2}$
 cooling lasers. The dashed line is the result of a Monte Carlo simulation of Sisyphus cooling of the molecular beam, with the amplitude of the cooling force as the only fit parameter \cite{li2018analysis}.}
\label{fig:BaH-Sysiphus}
\end{figure}

While laser deflection results presented in Sec. \ref{sec:Radiative-deflection} provide a valuable benchmark for the development of radiative slowing of BaH molecules, the data does not demonstrate a decrease  in the entropy of the molecular ensemble (as can be seen from the beam widths in Fig. \ref{fig:Deflection}(d)). To achieve a reduction in transverse velocity spread for the molecular beam, we establish a 1D standing light wave intersecting the molecular beam (Fig. \ref{fig:BaH-Sysiphus}(a)). Figure \ref{fig:BaH-Sysiphus}(b) demonstrates effective transverse temperature of the molecular beam as a function of the common detuning for the $\left(0,0\right)X^2\Sigma^+\rightarrow A^2\Pi_{1/2}$
 cooling laser with the repumping  $\left(0,1\right)X^2\Sigma^+\rightarrow B^2\Sigma^+$ laser fixed on resonance. We observe broadening of the molecular beam for red-detuned laser frequencies and narrowing for blue-detuned frequencies, consistent with Sisyphus laser heating and cooling of the ensemble, respectively \cite{kozyryev2017sisyphus,sheehy1990magnetic,shuman2010laser,lim2018laser}. 

To estimate the temperature of the beam we performed Monte Carlo simulations for various transverse temperature distributions, then selected the temperature that matches the observed expansion of the beam after a 5 mm aperture at the entrance to the interaction region. This relates the width of the unperturbed molecular beam as imaged on the camera to an effective transverse temperature. Sisyphus cooling applied over a 3 cm long interaction region reduced the effective transverse temperature by $\sim25\%$, from 20 mK to 15 mK, as shown in Fig. \ref{fig:BaH-Sysiphus}(b). We benchmark the strength of the Sisyphus cooling force by comparing a Monte Carlo simulation with only Doppler cooling to one with both Doppler and Sisyphus cooling. We find that achieved Sisyphus force is $\sim5$ times greater than Doppler, explaining why the Sisyphus cooling signature was more pronounced than pure Doppler cooling. 

\begin{figure}[h]
\centering
\includegraphics[width=.8 \textwidth]{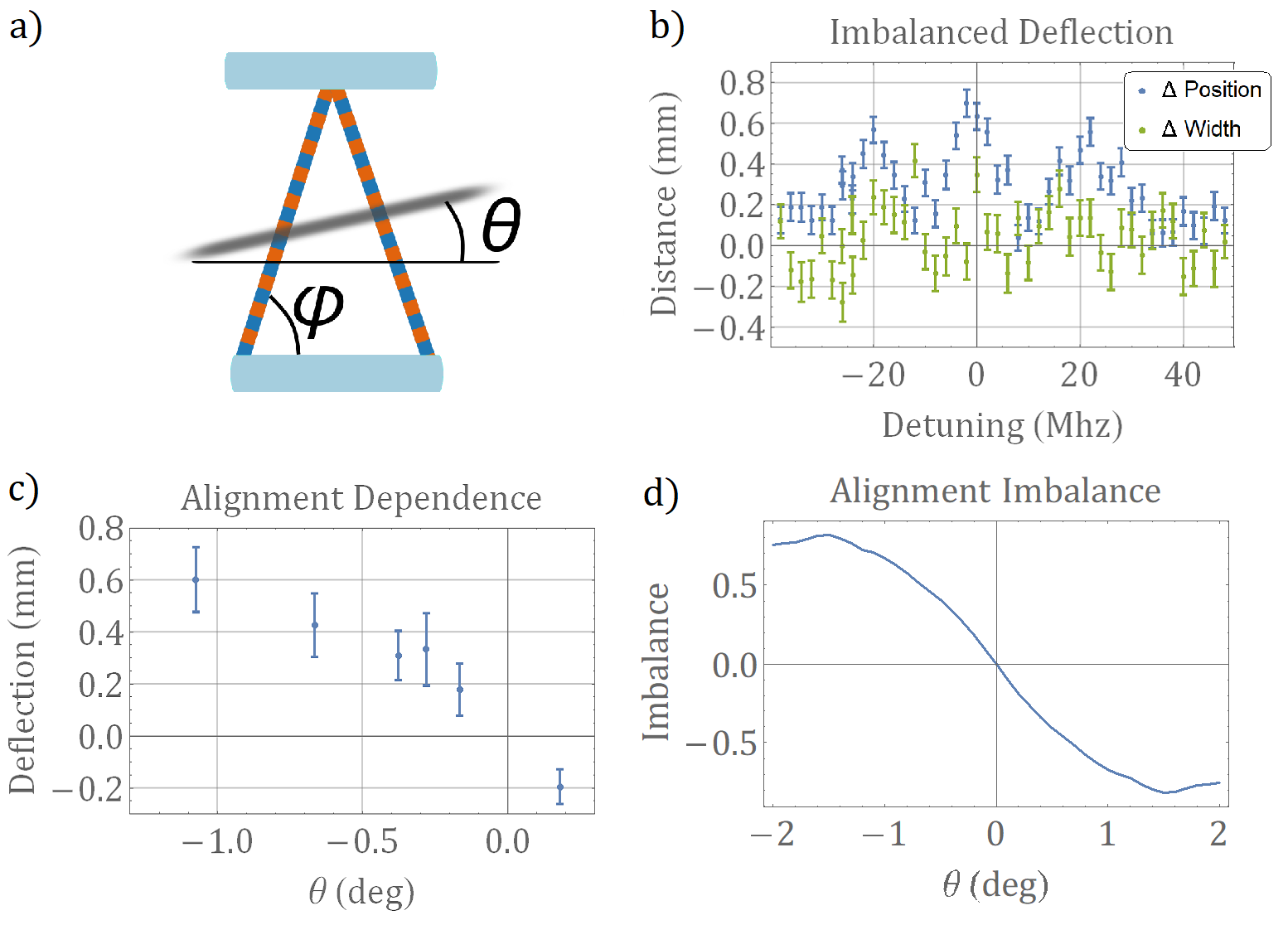}
\caption{a) A misalignment of the molecular beam relative to the multi-passed cooling laser. $\phi$ is the unavoidable angle present in the system to allow multiple passes, and $\theta$ is the misalignment between the parallel retroreflecting mirrors and the molecular beam. This misalignment results in different projections of the light propagation vectors onto the molecular beam forward velocity. b) Measured position and width of the molecular beam in a Doppler cooling configuration. The data shows a displacement due to a finite angular misalignment $\theta$. The three peaks are due to hyperfine structure of the cycling transition and the laser sidebands used to address them (see Appendix \ref{sec:BaH-rate-equations} for more details).  c)  As $\theta$ is varied in a controlled way, it can be experimentally minimized by zeroing the molecular beam deflection.  d) A  model accounting for the spread in the molecular beam's forward velocity can explain the observed beam deflection. Here, the imbalance is defined as the difference between the optimal force for each laser beam's direction divided by the maximum force.}
\label{fig:angle-imbalance}
\end{figure}

While attempting to observe Doppler cooling of the BaH beam by reducing the laser intensity and increasing the collimating aperture size to allow for a broader transverse velocity class to enter the cooling region, we have observed a systematic shift in the molecular beam position as shown in Fig. \ref{fig:angle-imbalance}. A small natural linewidth of the cooling transition ($\Gamma_{\rm{nat}}/2\pi\approx1.2$ MHz) leads to an acute dependence of the cycling rate on the alignment and detuning of the cooling laser relative to the molecular beam (Fig. \ref{fig:angle-imbalance}(a)).
The asymmetric Doppler shifts lead to a unidirectional deflection of the molecular beam in the cooling configuration (Fig. \ref{fig:angle-imbalance}(b)), where the direction of the deflection depends on the alignment angle as shown in the data in Fig. \ref{fig:angle-imbalance}(c). While such shifts were not important for transverse beam cooling of molecules with larger natural linewidth like SrF \cite{shuman2010laser} and SrOH \cite{kozyryev2017sisyphus}, a pronounced effect for misalignment of $<1^{\circ}$ observed in our work indicates that a careful geometry optimization will be required for performing precision spectroscopy for molecular beams of laser-coolable molecules with $\Gamma_{\rm{nat}}/2\pi\approx1$ MHz (e.g. TlF \cite{norrgard2017hyperfine} or TlCN \cite{Kudrin2019TlCN}). For a two level system, we can provide a simplified model of this effect as shown in Fig. \ref{fig:angle-imbalance}(d). This model uses a realistic Rabi rate for our experiment, and realistic distribution of forward velocity. Force imbalance ($\Delta F$) is based on the detuning dependent force from the left $F_l(\delta)$ (right $F_r(\delta)$) propagating laser beam: 
\begin{equation}
 \Delta F = \frac{max(F_l(\delta)) - max(F_r(\delta))}{max(F_l(\delta),F_r(\delta))}.
  \label{eq:Force-imbalance}
\end{equation}
We see that even small angular misalignment can lead to large imbalance in the maximum force pushing the beam to either direction.


\section{Sustained Cycling in a Longitudinal Configuration \label{sec:Slowing-config}}
In Sections \ref{sec:Optical-cycling} and \ref{sec:Radiative-deflection} we have confirmed optical cycling for $\sim80$ photons via depletion as well as transverse deflection of the molecular beam.  However, for slowing the beam to a magneto-optical trap (MOT) capture velocity over 30,000 photons need to be scattered in a longitudinal configuration. To demonstrate that we can maintain optical cycling for BaH in a slowing setup, we performed experiments with the cooling and repumping light counter-propagating against the molecular beam. The $\left(0,0\right)X^2\Sigma^+\rightarrow A^2\Pi_{1/2}$
 main cycling and  $\left(0,1\right)X^2\Sigma^+\rightarrow B^2\Sigma^+$ repuming light were both detuned to address a central velocity class of $140\, \rm{m/s}$ and broadened with a series of three EOMs (2, 5, and 15 MHz) until the observed laser spectrum was approximately flat on a scanning Fabry-Perot cavity, with a FWHM of $\sim70$ MHz. This so-called ``white light'' slowing \cite{barry2012laser,hemmerling2016laser} allows the laser light to be resonant with a majority of the forward velocity distribution of the cryogenic molecular beam.

\begin{figure}[h]
\centering
\includegraphics[width=.8 \textwidth]{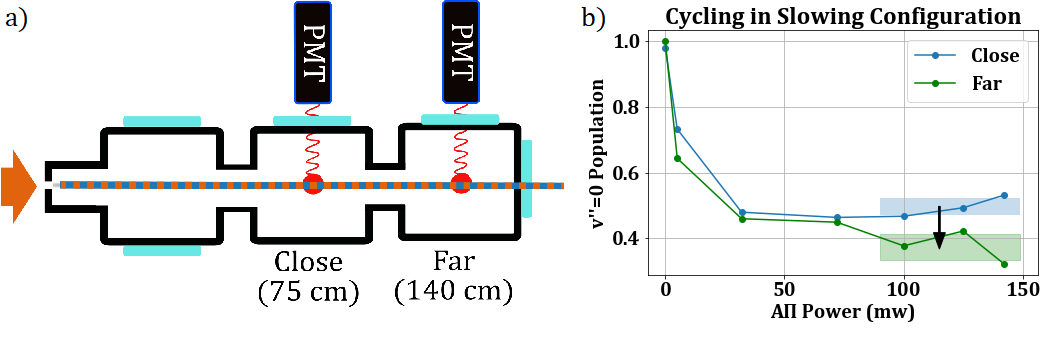}
\caption{a) Schematic of the experimental setup for the slowing experiment. The instantaneous $v''=0$ population is measured in two locations, 75 cm and 140 cm downstream from the molecular source.
b) Fractional population of $v''=0$ for a range of $\left(0,0\right)X^2\Sigma^+\rightarrow A^2\Pi_{1/2}$ light powers, with the $\left(0,1\right)X^2\Sigma^+\rightarrow B^2\Sigma^+$ power fixed at 100 mW. This measurement, shown for both the close and far locations, yields the average scattering rate in two ways, as described in the text. 
}
\label{fig:BaH-slowing}
\end{figure}

To study the scattering rate, we monitored the instantaneous $X^2\Sigma^+(v''=0)$ population at two points along the molecular beam propagation direction as a function of the A$^2\Pi_{1/2}$ light power, alternating the A$^2\Pi_{1/2}$ light on and off every shot. By taking the ratio of consecutive shots we reduce our sensitivity to molecular beam fluctuations and can isolate the effect of the A$^2\Pi_{1/2}$ light (Fig. \ref{fig:BaH-slowing}(a)). Because we detect molecules with the excitation to the E$^2\Pi_{1/2}$ state, while simultaneously cycling on the $\left(0,0\right)X^2\Sigma^+\leftrightarrow A^2\Pi_{1/2}$ transition, we can observe the instantaneous vibrational ground state fraction 75 cm  and 150 cm from the beam source; the presence of optical cycling will manifest as a reduction in the fraction of molecules residing in $v''=0$. While for laser powers below 90 mW the measured $v''=0$ population fraction is the same for both regions, for high A$^2\Pi_{1/2}$ light powers ($>$90 mW) we observe a $v''=0$ population of $49 \pm 3 \%$ in the near region and $37 \pm 4 \%$ in the far region (Fig. \ref{fig:BaH-slowing}(b)). Using ab initio calculations that utilized spectroscopically accurate molecular potentials for BaH \cite{moore2019assignment}, we attribute this population decrease to a combined loss into the $X^2 \Sigma^+(v''=2)$ excited vibrational state and $H^2\Delta_{3/2}$ metastable electronic state. We experimentally confirmed there is no dependence of the signal in the far region, on the presence of the E$^2\Pi_{1/2}$ laser in the close detection region. Based on the previous measurements and calculations of the BaH vibrational branching ratios \cite{moore2019assignment}, this $v''=0$ population reduction allows us to estimate $\sim4,500$ scattering events between the near and far regions or $\sim8,500$ total photon cycles. Given the time it takes the average molecule to reach the far detection region (11 ms), this gives a rate of $R_{\rm{scat}}\sim8\times10^{5}$ photons$/$s, consistent with what we measured using transverse beam deflection in Sec. \ref{sec:Radiative-deflection}. We see equal depletion for the full ToF beam profile, which indicates that we are able to maintain this high scattering rate for all forward velocities despite the additional complexity of achieving rapid photon cycing in the slowing configuration.

The data in Fig. \ref{fig:BaH-slowing}(b) suggests that the scattering rate bottleneck in this measurement is the repumping rate out of the $v''=1$ state. If the repumping light power was not limiting the overall scattering rate, there would be very little population in the $X^2\Sigma^+(v=1)$ state as we are decoupling the cycling and repumping lasers using two different electronic states. The main cycling $X-A$ laser light couples 12 ground state to 4 excited state sublevels, making 75\% an expected population fraction residing in $X^2\Sigma^+(v''=0)$ in a steady-state cycling configuration. 
Combining the $X^2\Sigma^+(v''=0)$ population measurement with the measured scattering rate from Sec. \ref{sec:Optical-cycling}, we can estimate state population of 50\%, 37.5\% and 12.5\% for the $X^2\Sigma^+(v=0)$, $X^2\Sigma^+(v=1)$ and $A^2\Pi_{1/2}(v=0)$ states, respectively. This reduction of the excited state population from the maximum attainable value of 25\% to 12.5\% leads to a reduction in the scattering rate to a value of $R_{\rm{scat}}\sim9\times10^{5}$ photons$/$s, consistent with the estimate based on the molecule loss to  $X^2\Sigma^+(v''=2)$ and $H^2\Delta_{3/2}$.

This high number of observed photon scattering events implies a reduction in the average beam velocity of $\sim25$ m/s, or $\sim15\%$. Unfortunately, the current apparatus is not capable of Doppler-sensitive forward velocity measurements since the narrow linewidth of the cooling transition and the large spread in forward velocities would reduce the signal-to-noise ratio by a factor of $\sim20$.
A planned upgrade using two-photon detection via a higher-lying electronic state, combined with a high-solid-angle detection system, should make this possible.

\section{Prospects for magneto-optical trapping of BaH \label{MOT-prospects}}
In order to characterize the feasibility for magneto-optical trapping for BaH molecules, we perform studies of confining forces using numerical solutions of the multilevel rate equation model following the framework presented in Ref. \cite{tarbutt2015magneto} and later used to model the MgF MOT properties \cite{xu2019three}. The simulation included $n_l=12$ magnetic sublevels of the $N''=1$ rotational manifold in the vibronic ground state (Fig. \ref{fig:level-structure}(b)), $n_u=4$ magnetic sublevels of the $J'=1/2$ manifold ($F'=0,\,1$) of the $v'=0$ vibrational level of the excited $A^2\Pi_{1/2}$ electronic state, and between three and six light frequency components from each direction. Because of the complex interplay between the ground and excited state $g$-factors as well as the specific nature of spacings between the hyperfine components in the ground vibronic state, a detailed numerical study is necessary in order to identify the optimal laser polarization structure and detunings\footnote{Numerical study of the dependence of the photon scattering rate on intensity and detuning of the $J''=3/2$ frequency components is provided in App. \ref{sec:OBE}.} \cite{tarbutt2015magneto}. Depending on whether the current in magnetic field coils used for MOT operation is static (DC) or alternating (AC), there are two types of molecular MOT operating regimes, correspondingly \cite{tarbutt2018laser,McCarron2018laser}. Moreover, in order to enhance the confining force in the DC MOT configuration, both ``blue'' and ``red'' detuned laser beam components can be applied resulting in a dual-frequency DC MOT \cite{tarbutt2015modeling}.

\begin{figure}[h]
\centering
\includegraphics[width=1 \textwidth]{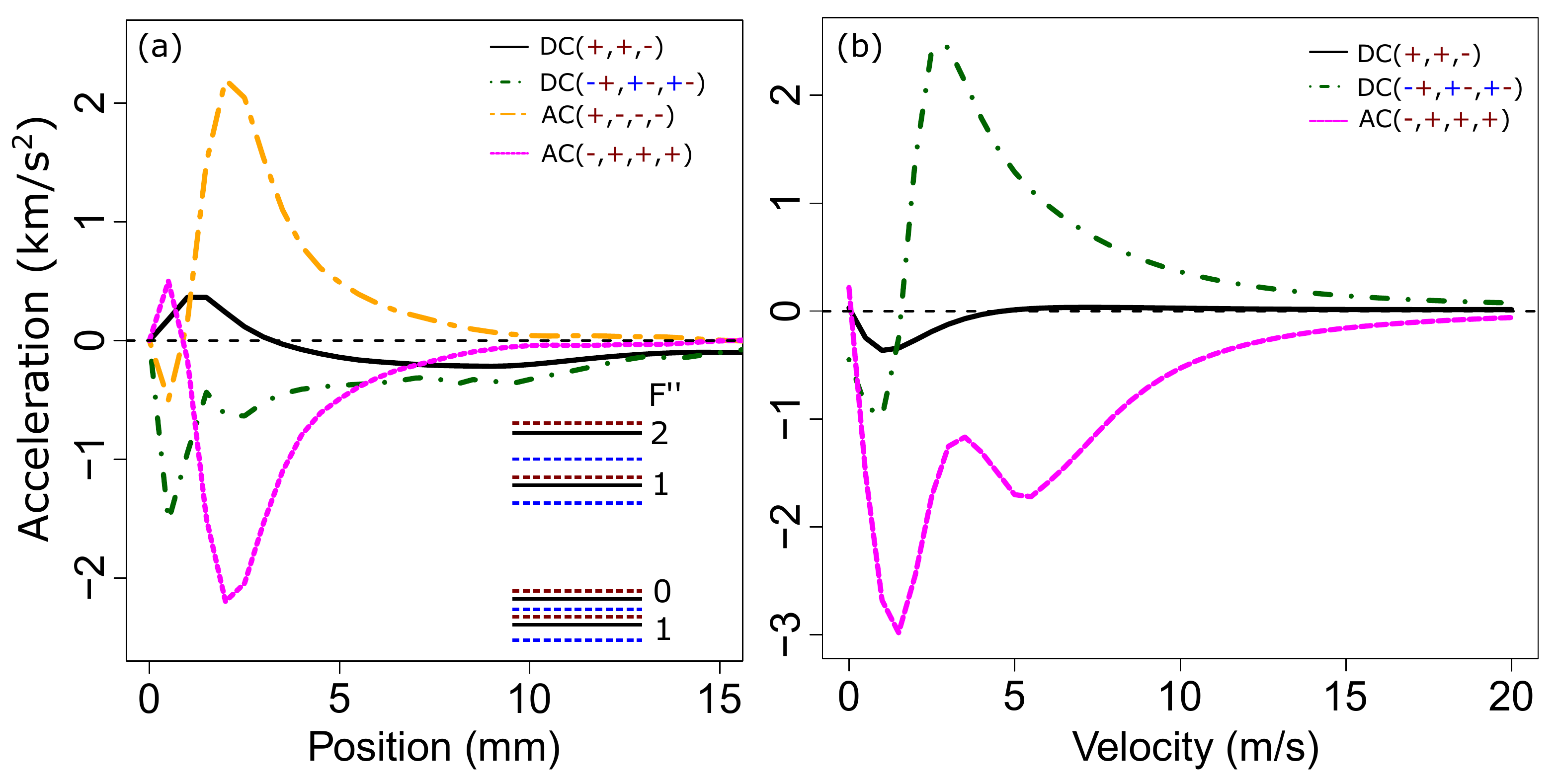}
\caption{Calculated BaH MOT confining (a) and cooling (b) characteristics for different polarization configurations for DC or AC MOT operation. The order of circular polarization labels in the legend starts with the lowest hyperfine substate, i.e. $(-,+,+,+)$ = ($-$ for $J''=1/2,\,F''=1$; $+$ for $J''=1/2,\,F''=0$; $+$ for $J''=3/2,\,F''=1$; $+$ for $J''=3/2,\,F''=2$) and $(+,+,-)$ = ($+$ for $J''=1/2,\,F''=1$; $+$ for $J''=3/2,\,F''=1$; $-$ for $J''=3/2,\,F''=2$). For the dual-frequency DC operation, two frequency components separated by approximately $3\,\Gamma_{\rm{sp}}$ were applied. Zeeman substructure of BaH is shown in Fig. \ref{fig:BaH-Zeeman} in the Appendix.}
\label{fig:BaH_MOT}
\end{figure}

Following earlier work \cite{tarbutt2015modeling,tarbutt2015magneto,xu2019three}, we present the strength of the achievable BaH magneto-optical trapping configurations by plotting molecular acceleration as a function of (a) the distance to the trap center for the $v=0$ velocity class, and (b) as a function of the velocity for a fixed position slightly displaced from the trap center ($d=0.1$ mm), in order to have a well defined quantization axis set by the direction of the magnetic field. 

Figure \ref{fig:BaH_MOT} provides a summary of the predicted acceleration profiles as a function of distance to the MOT center (for $v=0$ m/s) and molecular velocity (for $d=0.1$ mm) under experimental conditions approximating those in the experiments above ($P_{\rm{tot}}=200$ mW and 1-inch 1/e$^2$-diameter laser beams) and previously achieved in molecular MOT experiments (15 G/cm magnetic field gradient \cite{anderegg2017radio}). By comparing different MOT operation parameters (varying the number of laser frequencies and polarization settings) and configurations (AC vs DC vs DC dual-frequency), we conclude that the AC MOT has the highest potential for future trapping of BaH molecules providing both the highest peak deceleration ($\sim2$ km/s\textsuperscript{2}) and the largest velocity range affected by the MOT potential (up to $\sim 15$ m/s). We determine that the optimal polarization setting for the AC MOT is the same as that used for capturing CaF molecules \cite{anderegg2017radio}.

A unique property of BaH that distinguishes it from other molecules to which magneto-optical forces have been applied (SrF \cite{norrgard2016submillikelvin}, CaF \cite{anderegg2017radio}, YO \cite{collopy20183d,hummon2013YO} and CaOH \cite{baum2020CaOH}) is that a large excited state $g$-factor ($g_{\rm{eff}}\approx -0.51$ for the $A^2\Pi_{1/2}$ state \cite{iwata2017high}), arising from a strong mixing with the adjacent $B^2\Sigma^+$ electronic state, is approximately the same as that of the ground state ($g_{\rm{eff}}\approx +0.56$ for $J''=3/2$ \cite{iwata2017high}). Based on the model proposed in Ref. \cite{tarbutt2015magneto}, it  was anticipated that the DC configuration will lead to strong MOT confining forces for BaH molecules \cite{iwata2017high}. However, as can be seen from Fig. \ref{fig:BaH_MOT}(a), a complex interplay between the Zeeman shifts for the ground and excited magnetic sublevels contributes to a relatively weak confining force with an undesirable spatial structure and a small effective velocity range (Fig. \ref{fig:BaH_MOT}(b)). It was previously shown that for molecules with small Zeeman shifts in the excited state (like CaF \cite{tarbutt2015magneto} and MgF \cite{xu2019three}), the ``dual-frequency" contribution to the MOT forces far outweighs the confining effects arising from non-zero $g$-factors in the excited state. As can be seen from Fig. \ref{fig:BaH_MOT}, using the dual-frequency method outlined in Ref. \cite{tarbutt2015modeling} we can significantly improve the BaH MOT properties. However, in order to obtain a large velocity capture range, the use of the AC MOT configuration is necessary.  

\begin{figure}[h]
\centering
\includegraphics[width=0.6 \textwidth]{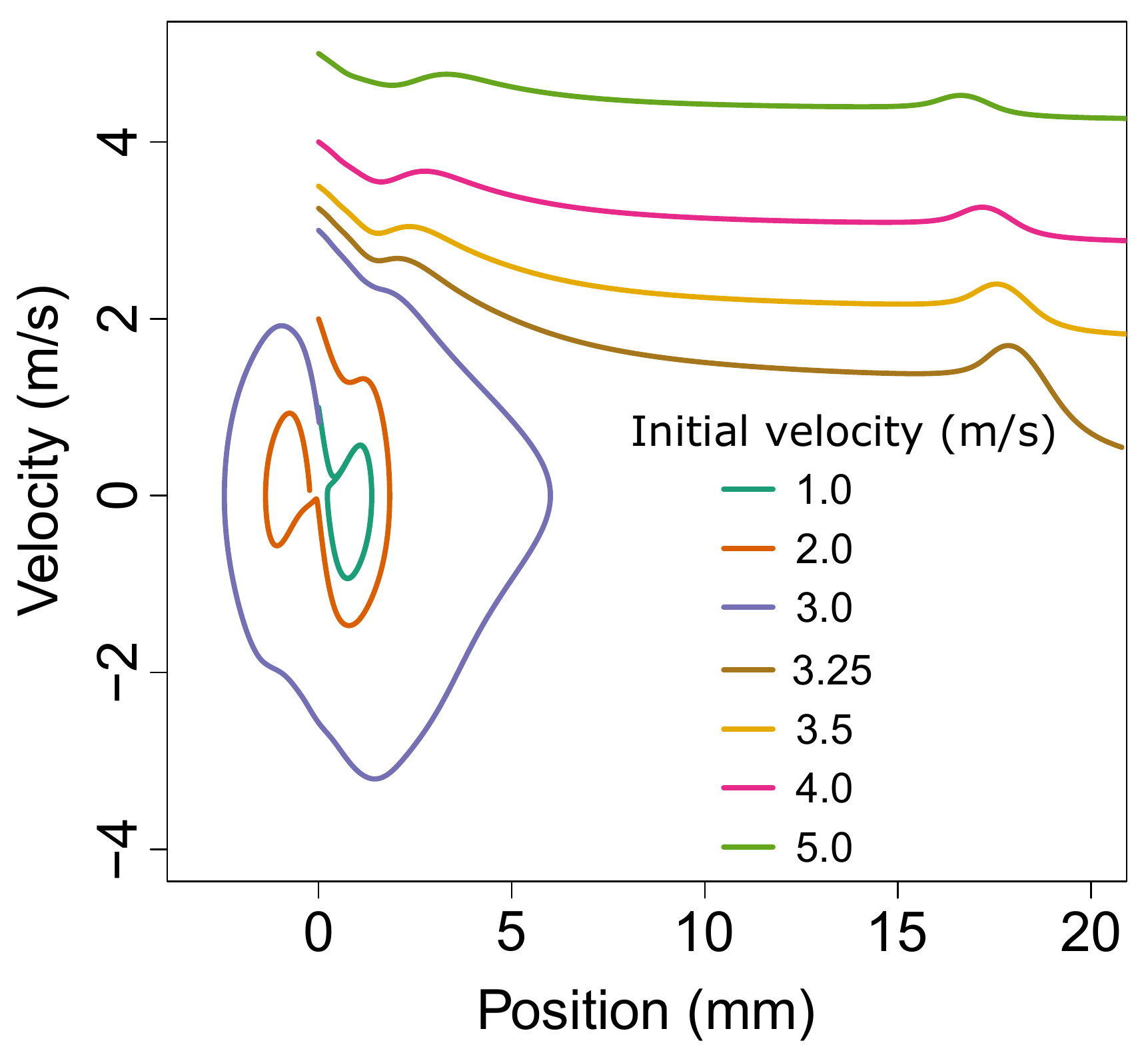}
\caption{Trajectories of BaH molecules inside a magneto-optical trapping potential, starting from the trap center with different initial velocities. From these trajectories, we estimate the MOT escape velocity to be $v_{\rm{esc}}\approx3$ m/s. The simulations are performed for the AC MOT configuration in Fig. \ref{fig:BaH_MOT}(b). The time evolution for the trapped and untrapped BaH trajectories is presented in the Appendix (Fig. \ref{fig:MOT-time-trajectories}).}
\label{fig:BaH_MOT_trajectories}
\end{figure}

Estimation of escape and capture velocities ($v_{\rm{esc}}$, $v_{\rm{cap}}$) are experimentally relevant ways to characterize the magneto-optical trapping potential.
3D MOT capture velocities have been measured for CaF ($v_{\rm{cap}}\approx 11$ m/s) \cite{williams2017characteristics} and SrF ($v_{\rm{cap}}\approx 5$ m/s) molecules \cite{steinecker2019sub} and provide useful benchmarks for our calculations. Previously it has been experimentally observed that an approximately linear relationship can be established between the MOT capture and escape velocities, $v_{\rm{cap}}=bv_{\rm{esc}}$, with a proportionality coefficient $b\gtrsim 1$ \cite{haw2012magneto,bagnato2000measuring}. Figure \ref{fig:BaH_MOT_trajectories} presents the simulated trajectories of BaH molecules that start at the geometric center of the MOT, for different initial velocities.
As shown in the plotted curves, we estimate  $v_{\rm{esc}}$ for the BaH AC MOT to be $\sim3$ m/s, leading to the MOT capture velocity $\sim3.5$ m/s.\footnote{Previously, the proportionality coefficient $b$ has been measured to be $1.2-1.4$ \cite{haw2012magneto,bagnato2000measuring,anwar2014revisiting}.} 

\section{Conclusion \label{discussion}}
Using a number of different methods, we have experimentally characterized optical cycling dynamics in BaH molecules. Specifically, we achieved a photon cycling rate of $8\times10^5$ s\textsuperscript{-1}, enabling radiative deflection and Sisyphus laser cooling of the BaH molecular beam. Furthermore, we provide evidence for uninterrupted scattering of up to $\sim 8,500$ photons in the longitudinal slowing configuration, limited by the length of the current vacuum chamber. These results validate the feasibility of laser cooling and trapping heavy molecules with relatively long-lived excited electronic states, despite previously unexplored technical challenges. In order to guide future work on laser slowing and magneto-optical trapping, we have carried out realistic estimations of MOT forces, pinned to the experimental benchmarks that we have already achieved. While our results demonstrate that a large Zeeman shift in the excited electronic state does not lead to a strong confining potential, an AC configuration of a magneto-optical trap has potential to capture BaH molecules with velocities $\lesssim3$ m/s. 

\section*{Acknowledgments}
 We would like to acknowledge helpful discussions with B. L. Augenbraun and Z. Lasner. Additionally, we thank Qi Sun for experimental assistance. This work was supported by the ONR Grant N00014-17-12246 and the W. M. Keck Foundation. RLM and KW gratefully acknowledge support by the NSF IGERT Grant No. DGE-1069240. IK has been supported by the Simons Junior Fellow Award.

\appendix
\section{Numerical solutions of the optical Bloch equations} 
\label{sec:OBE}

To obtain accurate theoretical estimates for the photon scattering rate we numerically solved the master equation for time evolution of the density matrix $\rho$ in the Lindblad form,
\begin{equation}
    \frac{d\rho(t)}{dt}=\mathcal{L}\rho(t),
\end{equation}
with $\mathcal{L}$ being the Lindblad superoperator of the form
\begin{equation}
    \mathcal{L}\rho(t)=-i\hbar[H,\rho]+\sum_{i=1}^{N^2-1}\gamma_i\left(C_i\rho C_i^{\dagger}-\frac{1}{2}\left\{C_i^{\dagger}C_i,\rho\right\}\right),
\end{equation}
where $C_i$ belong to a set of orthonormal operators with eigenvalues $\gamma_i$ and $N$ is number of states included \cite{koss, lind}. We used jump operators as our orthonormal set \cite{eit}, and in our case the dissipative part of the superoperator included only effects of spontaneous emission:  for a decay from state $\ket{i}$ to $\ket{f}$ with rate $\Gamma_{i\rightarrow f}$, we used $G_{i\rightarrow f}=\sqrt{\Gamma_{i\rightarrow f}}C_{i\rightarrow f}=\sqrt{\Gamma_{i\rightarrow f}}\ket{f}\bra{i}$.

\begin{figure}[h]
\centering
\includegraphics[scale=0.5]{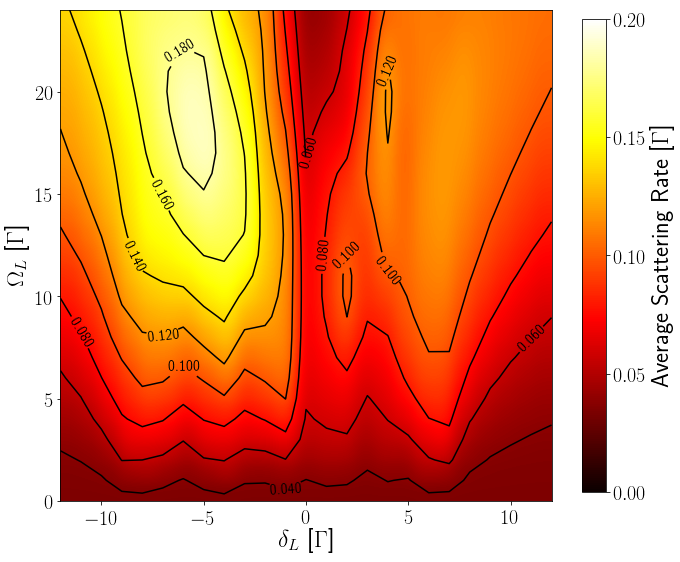}
\caption{Average scattering rate in the BaH
model as a function of the Rabi rate and detuning of the $J=3/2$ light. The solution was obtained for the $J=1/2$ light on resonance with Rabi rate $\Omega=13\,\Gamma$ and a background magnetic field $B=9\,\text{G}$. Both lasers were assumed to have linear polarizations perpendicular to each other, and the $J=1/2$ laser polarization was at an angle of 1 rad with respect to the quantization axis defined by the magnetic field. Magnetic field and laser polarization were assumed to lie in the same plane.}
\label{fig:Avg_scattering}
\end{figure}

In our calculations we included both spin-rotational manifolds $J=1/2$ and $J=3/2$ in the $v''=0$ vibrational state of the $X^2\Sigma^+$ ground electronic state with all the hyperfine levels and the $J'=1/2$ rotational manifold of the $A^2\Pi_{1/2}$ excited electronic state. We have also assumed no decay into higher vibrational states ($\mathcal{F}_{00}\approx 1$). Having set up the equations, we performed an optimization of the average scattering rate over the experimental interaction time $T$,
\begin{equation}
\overline{\Gamma}=\sum_{i}\frac{1}{T}\int_0^T\rho_{e_ie_i}(t)\Gamma dt,
\end{equation}
\noindent where the sum is over all excited states decaying with the same rate $\Gamma$. The optimization was performed with respect to the Rabi rate for the $\ket{X^2\Sigma^+;J=1/2}$ to $\ket{A^2\Pi_{1/2};J=1/2}$ transition, Rabi rate for the $\ket{X^2\Sigma^+;J=3/2}$ to $\ket{A^2\Pi_{1/2};J=1/2}$ transition, detunings of both transitions, polarization of the light fields, and the background magnetic field responsible for dark state remixing. Given the experimental constraints, we found the maximum achievable average scattering rate of $\overline{\Gamma}\approx \Gamma/5.21$, which agrees well with the highest scattering rate we achieved in the experiment (Sec. \ref{sec:Radiative-deflection}).

In Fig. \ref{fig:Avg_scattering} we show the average scattering rate obtained in the simulations as a function of the Rabi rate and detuning
of the $J=3/2$ laser. We observe that the scattering rate is highest for relatively large Rabi rates, which can be expected since the excitation rates have to match remixing rates in order to reach optimal values \cite{Berkeland2002}. We also see that, because of the nature of our coupling scheme where we effectively create a $\Lambda$-type system with many more ground states than excited states, having both lasers on resonance is detrimental to achieving high scattering rates. 

\section{BaH Zeeman structure in the ground vibronic state \label{sec:BaH-Zeeman}}
\begin{figure}[h]
\centering
\includegraphics[width=1 \textwidth]{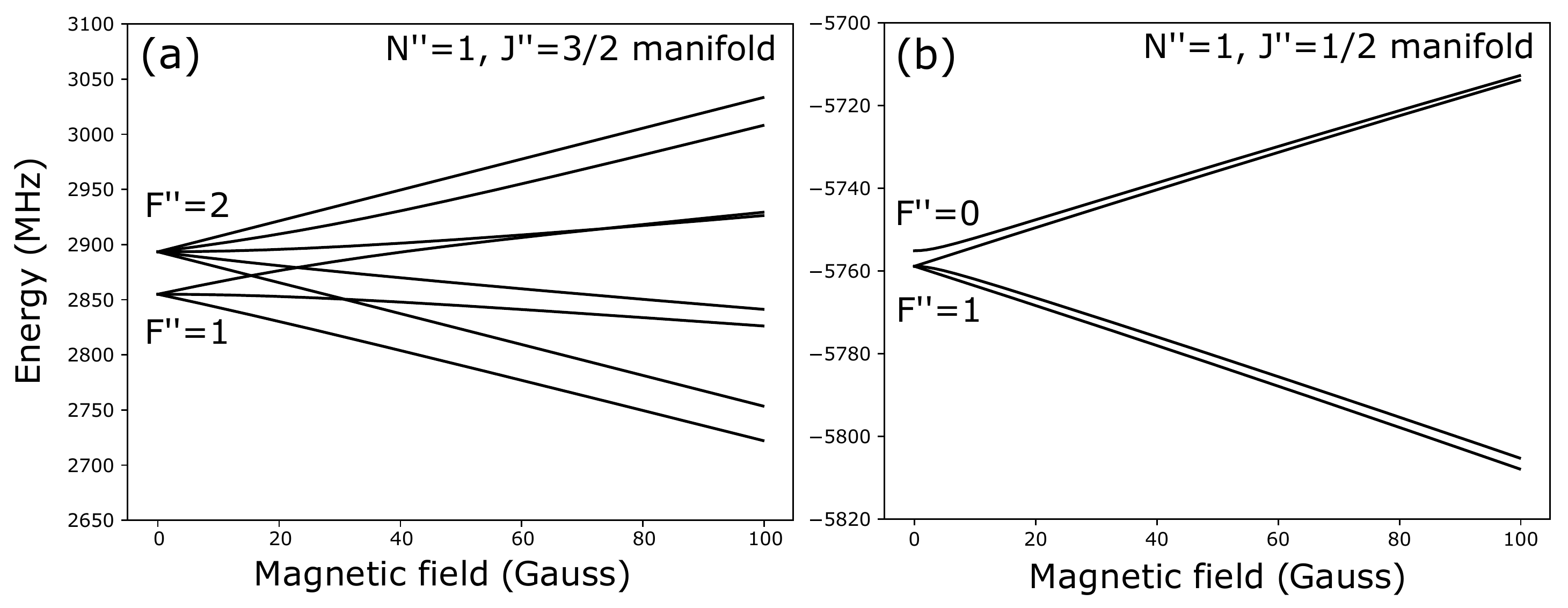}
\caption{Calculated Zeeman shifts for the $\ket{X,\,N''=1}$ manifold of BaH in magnetic fields up to 100 G for the $J''=3/2$ (a) and $J''=1/2$ (b) manifolds.
To model the magneto-optical trapping process we use $g_F$ fitted values for the $J''=3/2$ states and $g_J$ values for the $J''=1/2$ states.}
\label{fig:BaH-Zeeman}
\end{figure}

In order to develop a proper theoretical model of the magneto-optical trapping forces for molecules, an accurate approximation of the molecular energy shifts with applied magnetic fields (Zeeman shifts) is necessary. Zeeman shift measurements for BaH have been previously reported \cite{iwata2017high} in the intermediate magnetic field range ($50-100$ G). Using the extracted hyperfine and spin-rotation parameters, we can estimate the Zeeman energy levels more generally using the following Hamiltonian,
\begin{equation}
\label{eq:Hz}
    H = \gamma_{\rm{SR}} \mathbf{N}\cdot \mathbf{S} + b\mathbf{I}\cdot \mathbf{S} + cI_ZS_Z + \mu _{\rm{B}}g_S\mathbf{B}\cdot \mathbf{S},
\end{equation}
with the spin-rotation constant $\gamma_{\rm{SR}}$ and hyperfine constants $b,\,c$ taken from our previous measurements \cite{iwata2017high}. Figure \ref{fig:BaH-Zeeman} shows the results of the Hamiltonian in Eq. \ref{eq:Hz} diagonalized for a range of fields between 0 and 100 G. As can be seen from Fig. \ref{fig:BaH-Zeeman}(a), a linear approximation for the $m_F$ Zeeman sublevels is valid for fields up to $\sim30$ G for the $\ket{N''=1,\,J''=3/2}$ manifold, while because of the small hyperfine splitting of the $\ket{N''=1,\,J''=1/2}$ sublevels, we use a linear approximation for the $m_J$ Zeeman sublevels.

\section{Rate equation modeling of the optimal scattering rate \label{sec:BaH-rate-equations}}

\begin{figure}[h]
\centering
\includegraphics[width=1 \textwidth]{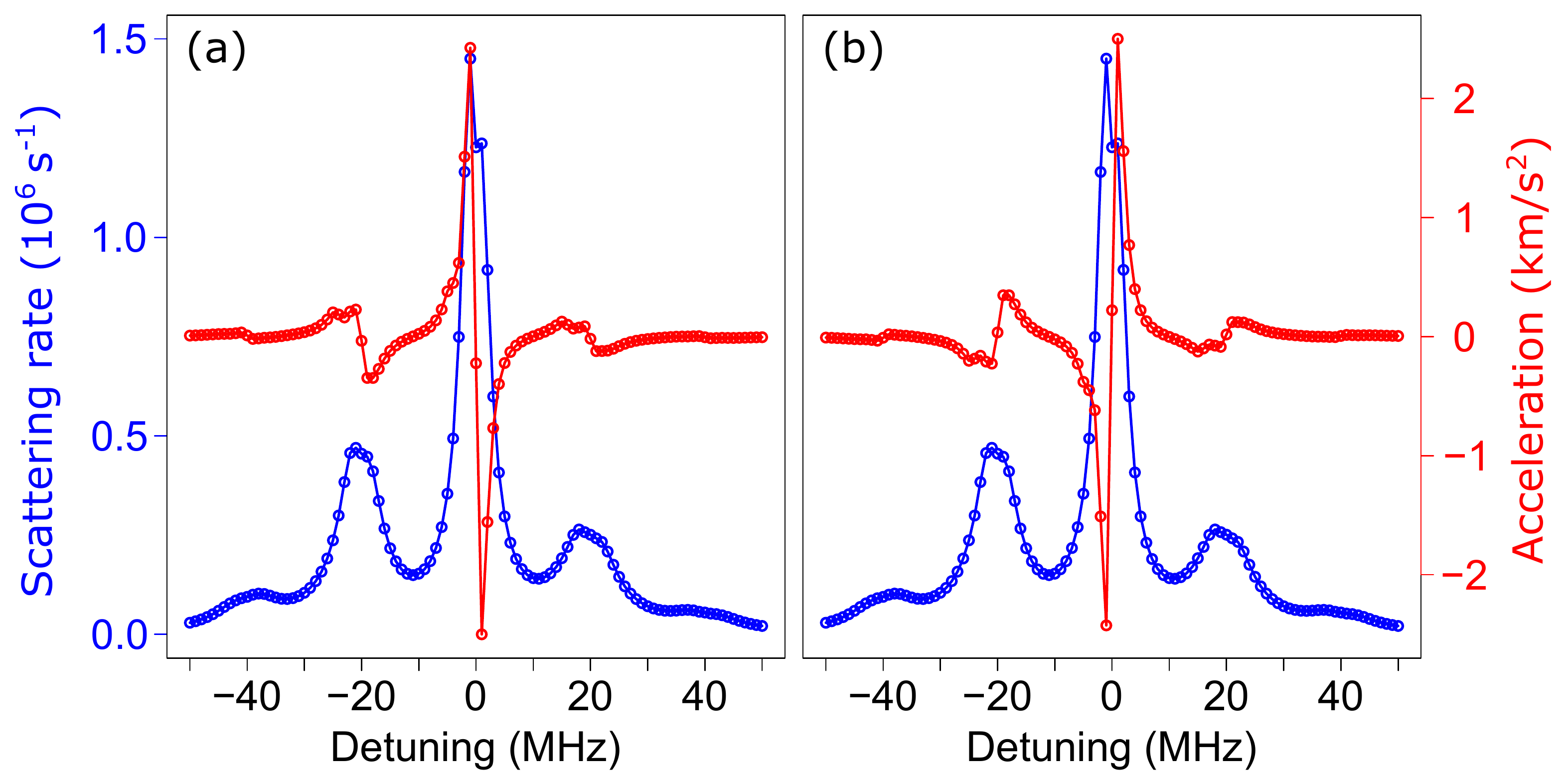}
\caption{Predicted scattering rate and acceleration with the 20 MHz sidebands generated by the EOM for two different transverse velocity classes: (a) $v_{\rm{trans}}=-1$ m/s and (b) $v_{\rm{trans}}=1$ m/s. The rate equation model was solved under the experimental parameters appropriate for the data presented in Fig. \ref{fig:angle-imbalance} with $\phi = 89^{\circ}$ and $\theta = 1^{\circ}$. 1 MHz frequency steps were used in the simulation which is comparable to the laser stability in the experiment.}
\label{fig:Predicted-cycling}
\end{figure}

In order to better understand the laser cooling process accompanying the observed beam deflection signature described in Sec. \ref{sec:Laser-cooling}, we solved the rate equation model with the three EOM sidebands spaced by 20 MHz. Results presented in Fig. \ref{fig:Predicted-cycling} can be compared to experimental observations in Fig. \ref{fig:angle-imbalance}(b). Notice the qualitative agreement between the blue curves in both figures, as we observe peak beam center shift at detuning of $0,\,\pm 20$ MHz. We also detected beam width change around the -20 MHz detuning, consistent with the calculated acceleration curve. However, we do not observe consistent width change arising from the Doppler forces (modelled by the rate equations) around zero detuning, perhaps due to cancelling out between the Doppler and Sisyphus effects.

\section{Simulating time dynamics of trapped BaH molecules\label{sec:BaH-MOT-profiles}}

\begin{figure}[h]
\centering
\includegraphics[width=1 \textwidth]{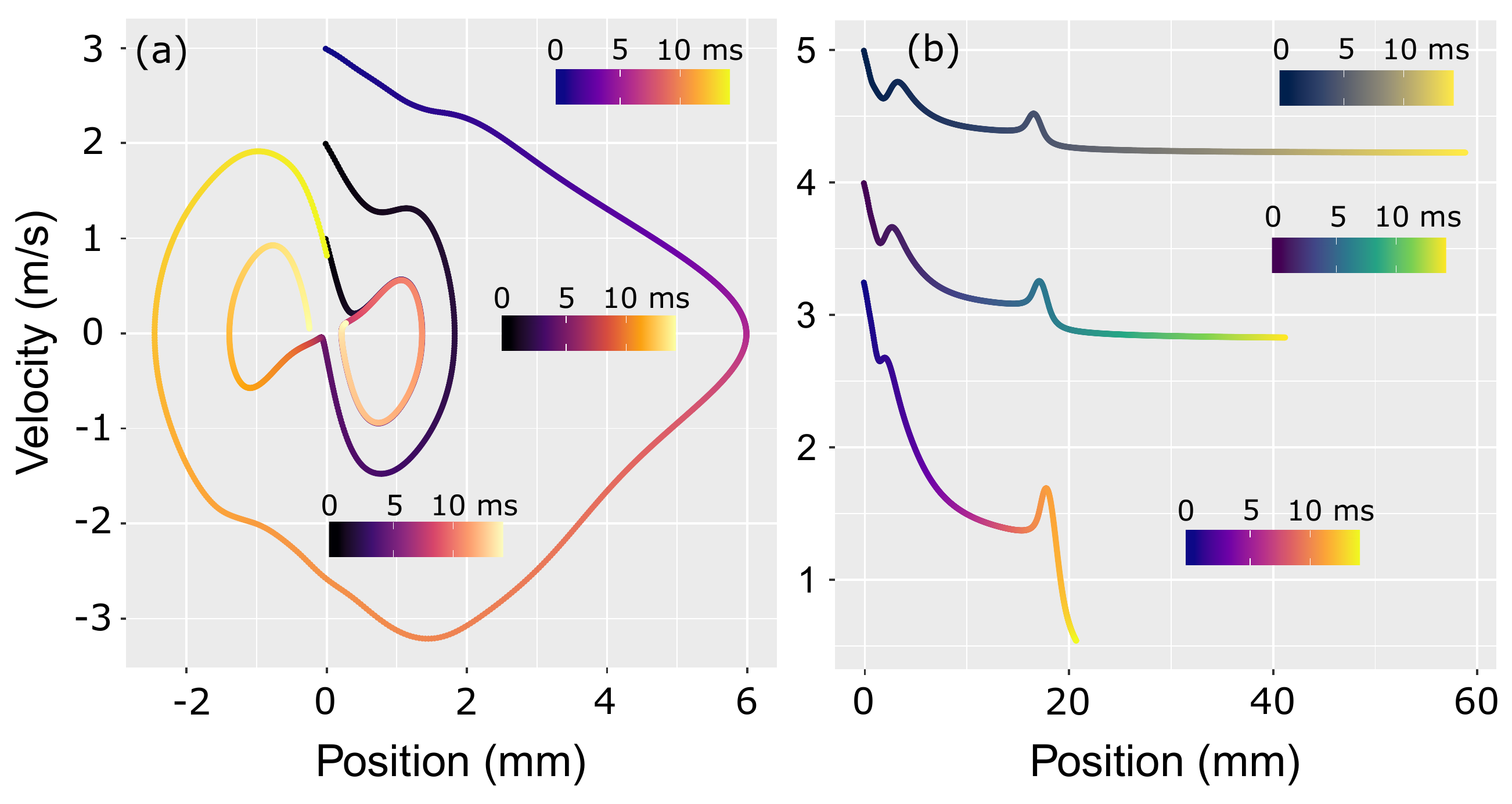}
\caption{Simulated time evolution of BaH molecules inside a magneto-optical trapping potential for different initial velocities starting at the trap center for (a) trapped and (b) untrapped trajectories. Color gradient scale provides time information not available in Fig. \ref{fig:BaH_MOT_trajectories}.}
\label{fig:MOT-time-trajectories}
\end{figure}

Large vibrational spacing for diatomic monohydrides (MH) compared to monofluorides (MF) and monohydroxides (MOH) will lead to enhanced vibrational decay from the $v''=1$ levels populated during the MOT loading process. Previous accurate ab initio calculations for BaH predict spontaneous vibrational lifetime for $v''=1,\,N''=1$ state of $\tau_{\rm{vib}}=12.6$ ms \cite{moore2019assignment}. In order to understand the time evolution of BaH trajectories inside a magneto-optical potential for different initial velocities, we plotted both trapped and untrapped BaH trajectories in Fig. \ref{fig:MOT-time-trajectories} with the time information provided as a color gradient. In order to minimize the loss of molecules to dark rotational sublevels due to spontaneous vibrational decay, one would ideally transfer BaH molecules from a MOT into a conservative magnetic \cite{mccarron2018magnetic,williams2018magnetic} or optical \cite{anderegg2018dipole} trap within the first $\sim 10$ ms of capturing them in a MOT.

%

\end{document}